\documentstyle[12pt]{article}
\include{epsf}
\def\lo{\langle 0 |}
\def\ro{ | 0 \rangle }

\def\gmmu{\gamma _{\mu}}
\def\gmnu{\gamma_{\nu}}
\def\atop{ \frac{ \alpha_{s}}{4 \pi} G_{\mu \nu}
 \tilde{G}_{\mu \nu} }
\def\gcub{ g^3 f^{abc} G_{\mu \nu}^a \tilde{G}_{\nu 
\alpha}^b G_{\alpha \mu}^c }
\def\fc{ f_{\eta'}^{(c)} }
\def\gmf{\gamma _{5}}

\def\la{\langle }
\def\ra{ \rangle }
\def\el{ \langle \eta'| }
\def\er{ | \eta' \rangle}
\def\pdir{ P \! \! \! \! / }

\newcommand{\beq}{\begin{equation}}
\newcommand{\eeq}{\end{equation}}
\newcommand{\bea}{\begin{eqnarray}}
\newcommand{\eea}{\end{eqnarray}}

\begin{document}
                                        \begin{titlepage}
\begin{flushright}
hep-ph/9704412
\end{flushright}
\vskip1.8cm
\begin{center}
{\LARGE
  $ B \rightarrow K \eta' $ decay as unique probe of 
$ \eta' $ meson }   
            
\vskip1.5cm
 {\Large Igor~Halperin} 
and 
{\Large Ariel~Zhitnitsky}
\vskip0.2cm
        Physics and Astronomy Department \\
        University of British Columbia \\
        6224 Agricultural Road, Vancouver, BC V6T 1Z1, Canada \\  
     {\small e-mail: 
higor@physics.ubc.ca \\
arz@physics.ubc.ca }\\
\vskip1.5cm
{\Large Abstract:\\}
\end{center}
\parbox[t]{\textwidth}{  
A theory of the $ B \rightarrow K \eta' $ decay is 
proposed. It is based on the Cabbibo favored $ b \rightarrow
\bar{c} c s $ process followed by a direct materialization of 
the $ \bar{c} c $ pair into the $ \eta'$. This mechanism works 
due to a non-valence Zweig rule violating $c$-quark component 
of the $\eta'$, which is unique to its very special nature. 
This non-perturbative ``intrinsic charm" content of the $\eta'$ 
is evaluated using the Operator Product Expansion and QCD low 
energy theorems. Our results are consistent with an unexpectedly 
large $ Br( B \rightarrow K \eta') \simeq 7.8 \cdot 10^{-5} $  
recently announced by CLEO.
 }

\vspace{1.0cm}

                                                \end{titlepage}

\section{Introduction}

This paper suggests a theory of the 
$ B \rightarrow K \eta' $ decay which may shed a new light
on properties of the $ \eta' $ meson. 
Our study is motivated by recent results 
of the CLEO collaboration \cite{CLEO} which has
announced 
an unexpectedly large branching ratio
\beq
\label{Kim}
Br(B \rightarrow K \eta') = (7.8_{-2.2}^{+
2.7} \pm 1.0) \cdot 10^{-5}
\eeq
Little thought is needed to realize that 
this number is in severe contradiction 
with a standard view of the process at the 
quark level as a decay of the $b$-quark into 
the light quarks which could be suggested as soon as the  
 $ \eta' $ is usually considered to be a SU(3) singlet
meson made of the $u-$, $d-$ and $s-$quarks (see 
Sect.2 for more detail). This result may not seem 
too surprising if one remembers the well known fact 
that the quark content of the $ \eta' $ is 
undistinguishable from the gluon one due to the 
axial anomaly. One the other hand, in the 
weak decay the $b$-quark proceeds more strongly
to the $ \bar{c} c s $ system due to the Cabbibo 
enhancement of the latter in comparison to the $ \bar{
u} u s $ state. Since a pair of $c$-quarks can 
easily convert to gluons, one can suggest the 
following scenario of the $ B \rightarrow K \eta' $
decay. The $b$-quark proceeds into the $s$-quark
and $c$-quark pair, while the latter
directly materializes
into the $ \eta' $ via a non-valence 
``intrinsic charm" $c$-quark
component of the $\eta'$ which exists due to 
virtual $ \bar{c} c \leftrightarrow gluons$ 
transitions. 
An immediate objection to this proposal which 
can come to one's mind is that this process 
is expected to bring a very small contribution 
to the decay width as soon as it obviously  
violates the Zweig rule. We will argue that 
though the scenario we suggest is indeed 
Zweig rule-violating, it nevertheless can 
explain the data. The reason is that we actually
deal here with a situation where the Zweig rule 
itself is badly broken down. As will be discussed in 
detail in Sect.4, both regularities and sources 
of breaking down the Zweig rule are nowadays well
classified and studied. In particular, it is 
100 \% violated for pseudoscalar mesons including
the $ \eta'$. In effect, we find that 
an extent to which the Zweig rule is 
broken down in the problem at hand suffices to  
reconcile the theory with the data (\ref{Kim}).
The uniqueness of $ \eta'$ is in both a possibility 
to evaluate this effect and its very large magnitude.
{\bf The decay $ B \rightarrow K \eta' $ serves as 
a probe of the ``intrinsic charm" content of the $
\eta'$. }
On the contrary to what it may sound, the
mechanism of violating the Zweig rule in the 
$ \eta' $ is purely non-perturbative.  
To be honest, we have to note that an 
accuracy of our result is rather low, of the
order of factor two in the amplitude. 
It is important, however, that a main source
of uncertainty in our approach is well 
localized and related to a poor knowledge 
of a particular vacuum condensate.  
Therefore the theoretical precision can
be considerably improved in the future.
We should stress that in contrast to a recent 
proposal \cite{Soni} on importance of the axial
 anomaly in the 
closely related inclusive $ B \rightarrow \eta' X_s $ 
decay, it does not play any role in our mechanism.
On the contrary, the anomaly is exactly cancelled in 
the Operator Product Expansion (OPE) in powers of 
$ m_c^{-1} $ for a $c$-quark bilinear operator
(see Eqs.(\ref{11},\ref{12}) below), which is a 
starting point of our approach to the problem. In
a sense, we therefore deal with a ``post-anomalous"
effect which, of course, is suppressed by the parameter
$ 1/m_c^2 $. However, in the real world $ m_c \simeq 1.25 \; 
GeV $ is not far from a hadronic $ \sim 1 \; GeV $ scale
and, as will be shown below, an effect of the charmed 
loop is very large numerically.
 On the other hand, methods applied in our 
study are in close parallelism with those developed
earlier in a study of the famous $ U(1) $ problem
(whose key ingredient is just the axial anomaly),
and will be explained in the course of our presentation.

Our strategy consists of a few steps. We start 
in Sect.2 with the standard approach 
to the $  B \rightarrow
 K \eta' $ decay and demonstrate that its 
prediction is about two orders of magnitude 
smaller than the experimental number (\ref{Kim}).
We then propose in Sect.3 an alternative gluon
mechanism and explain
our method for calculation of a crucial 
quantity of our consideration which is the matrix 
element $ \lo \bar{c} \gmmu \gmf c \er $. 
Using the data (\ref{Kim}) as an input, we 
calculate an ``experimental" value of this
matrix element. To calculate the same 
quantity theoretically, we first
reduce it by using the OPE to the matrix element 
of a pseudo-scalar three gluon 
operator $ \lo G \tilde{G} G \er $. The latter object is   
further related to a 
particular correlation function of gluon 
currents extending ideas originally suggested 
by Witten \cite{Witten} 
and Veneziano \cite{Ven} in their approach to the $ U(1) $ problem.
This correlation function is next calculated in Sect.4
in terms of a vacuum expectation value of the three gluon 
operator $ \la g^3 G^3 \ra $ by  
 using QCD low energy theorems. We also 
discuss there physics responsible 
for breaking down the Zweig rule.
In Sect.5 we estimate the latter vacuum condensate
and finally obtain a theoretical prediction for the 
matrix element of interest  $ \lo \bar{c} \gmmu \gmf 
c \er $. We compare this number with an ``experimental"
value found from the observed data (\ref{Kim}) and 
find a satisfactory agreement between them. 
This demonstrates that the gluon  
mechanism indeed explains the data with a reservation for 
uncertainty of our results. A final 
Sect.6 contains our conclusions.    
      
\section{The standard approach to $ B \rightarrow
 K \eta' $}

In this section we estimate a width of  the 
$ B \rightarrow K \eta' $ decay assuming that the
 $ \eta' $ meson 
is made exclusively of light quarks. In this case
 the relevant terms 
in the effective non-leptonic Hamiltonian are 
\beq
\label{1}
H_{\delta B = 1} = \frac{G_{F}}{\sqrt{2}} \left[
V_{ub} V_{us}^{*} ( c_{1} O_{1} + c_{2} O_{2} ) 
- V_{tb} V_{ts}^{*} \sum_{i=3}^{12} c_{i} O_{i} 
\right] + h.c.
\eeq
Here $ O_{i} $ are defined as (we use
the notations $ L_{\mu} = 
\gmmu (
1-\gmf) , R_{\mu} = \gmmu ( 1+ \gmf) $ ) 
\bea
\label{2}
O_{1} = \bar{s}^i L_{\mu} u^j \bar{u}^j L_{\mu} b^i 
& , & O_{2} = \bar{s} L_{\mu} u \bar{u} L_{\mu} b
\nonumber \\
O_{3(5)} =  \bar{s} L_{\mu} b \sum_{q} \bar{q}
 L_{\mu}(R_{\mu}) q
& , & O_{4(6)} =  \bar{s}^i L_{\mu} b^j \sum_{q} 
\bar{q}^j
 L_{\mu}(R_{\mu}) q^i  \nonumber \\
O_{7(9)} = \frac{3}{2}  \bar{s} L_{\mu} b 
\sum_{q} e_{q} \bar{q} R_{\mu} (L_{\mu}) q
&,& O_{8(10)} =  \frac{3}{2}  \bar{s}^i L_{\mu} b^j 
\sum_{q} e_{q} \bar{q}^j R_{\mu} (L_{\mu}) q^i  \\
O_{11} = \frac{g}{32 \pi^2} m_{b} \bar{s} ( 1+ \gmf)
\sigma G 
 b & , & O_{12} = \frac{e}{32 \pi^2} 
m_{b} \bar{s} (1+ \gmf) \sigma F b  \nonumber 
\eea
where $ i , j $ are the color indices and $ q $ is
 any of the 
 the $ u , d,s,c $ quarks. $ G_{\mu \nu} =  
 G_{\mu \nu}^a t^a $ and 
$ F_{\mu \nu} $ are the gluon and photon field strength
 tensors.
$ O_1 $ and $ O_2 $ are the tree level operators,
 while $ O_{3-6} $
and $ O_{7-10} $ are the gluon and electroweak penguin
 operators,
respectively. $ O_{11,12} $ are the magnetic penguins. 
The Wilson
coefficients $ c_{i} = c_{i}(\mu) $ depend on the 
renormalization 
scale $ \mu $ and to the next-to-leading order 
\cite{Buras}
( for $ \alpha_{s}(m_Z) = 0.118, \alpha_{em}(m_Z) = 
1/128 , m_{t}
= 176 \; GeV, \mu \simeq 5 \; GeV $ ) are given by the
set 
\cite{Buras,Desh,Ciu}
\bea 
\label{3}
c_1 = -0.3125 &,& c_2 = 1.1502 \; , \; c_3 = 0.0174
 \; , \; 
c_4 = - 0.0373 \; , \nonumber \\
c_5 = 0.0104 &,& c_6 = - 0.0459 \; , \; c_7 = 1.398 
\cdot 10^{-5}
\; , \\
c_8 = 3.919 \cdot 10^{-4} &,& c_9 = - 0.0103 \; , \; 
c_{10} = 1.987 \cdot 10^{-3} \nonumber \\
c_{11} = -0.299 &, & c_{12} = -0.634  \nonumber 
\eea
Introducing the transition form factor 
\beq
\label{4} 
 \la K(q) | \bar{s} \gmmu b | B(p+q) \ra = 2 q_{\mu}
f_{+} (m_{\eta'}^2) + p_{\mu} \left( f_{+} 
(m_{\eta'}^2) + 
 f_{-} (m_{\eta'}^2) \right)
\eeq
 and the $ \eta'$ residue \cite{NSVZ} (the chiral
 limit $
m_q = 0 $ is implied)
\beq
\label{5}
\el \bar{q} \gmmu \gmf q \ro = - i 
\frac{1}{\sqrt{3}} f_{\eta'}  p_{\mu} \; , \; 
\frac{1}{\sqrt{3}} f_{\eta'} = (0.5 - 0.8) \cdot 
\frac{f_{\pi}}{
\sqrt{3}} \simeq 
0.04 \; GeV
\eeq
and neglecting for the 
moment $ m_{\eta'}, m_{K} $ in comparison to $ m_B $,
the magnetic penguins $ O_{11}, O_{12} $ and
  $ O(1/m_b , 
1/N ) $ terms in the factorized matrix elements 
of penguin operators, 
we obtain the following  estimate
 for the amplitude of interest (here $ N $ stands for the 
number of colors)
\bea
\label{6}
\la K \eta'| H_{W}| B \ra \simeq \frac{G_{F}}{\sqrt{2}}
 2 i (pq)
f_{+}(m_{\eta'}^2)  \frac{f_{\eta'}}{
\sqrt{3}}\left[ V_{ub} V_{us}^* (c_1 + 
\frac{c_2}{N}) \right.
\nonumber \\ 
\left. - V_{tb}V_{ts}^* ( 3 c_3 + c_4 - 3 c_5 + \frac{3}{2}
 e_s c_{10} ) \right] 
\eea
where, in particular, we have omitted left-right penguin 
contributions, which are suppressed by $ 1/m_b $. In our opinion,
this procedure is much better than an alternative one, where only 
a subset of $ 1/m_b $ corrections is retained.  
Furthermore, it is well known that the factorization  does
not work in non-leptonic B-decays. Effects due to a 
non-factorizability are usually taken into account in a
 phenomenological manner by the substitution $ c_1 + c_2 /N 
\rightarrow a_1 $ with $ a_1 \simeq 0.25 $ obtained by a global
fit of the data on non-leptonic B-decays \cite{B-decays}. 
Using this number in (\ref{6}), we end up with\footnote{We 
disagree with \cite{Chau} where much larger $ 
Br( B \rightarrow K \eta') \simeq 3 \cdot 10^{-5}
$ was  proposed. In our opinion, this large width came as a result
of an incorrect assignment of absorptive parts
to matrix elements of penguin operators, which, by definition of the 
OPE, are not there. 
In particular, it follows (see p. 2187) from the formulas 
given in \cite{Chau} that this decay width becomes infinite (?)  
in the chiral SU(3) limit. In fact, at the level of penguin 
contributions
the decays $ B \rightarrow K \eta' $ and $ B \rightarrow K \phi $ 
are just identical and, assuming that factorization
works reasonably well, it is simply 
impossible to obtain anything substantially different from our 
estimate (\ref{8}).}
\beq
\label{7}
Br( B \rightarrow K \eta') \simeq 1 \cdot 10^{-7}
\eeq
which is by two order of magnitude smaller than the experimental
result \cite{CLEO}. It is easy to see that this small value is 
a consequence of a small residue of the $ \eta' $
supplemented with the Cabbibo suppression of the 
$ b \rightarrow u $ transition. An account of penguin 
contributions, as can be seen from (\ref{6}), does not help much.
Indeed, neglecting for simplicity the tree level $ b \rightarrow
u $ transition, we can obtain an 
estimate for the ratio of the decay width of the process of 
interest to the width of the decay $ B \rightarrow K \phi $:
\beq
\label{8}
\frac{\Gamma(B \rightarrow K \eta')}{\Gamma(B \rightarrow 
K \phi)} \simeq \frac{ | \el \bar{s} \gmmu \gmf s \ro \la 
K | \bar{s} \gmmu b | B \ra |^2}{  | \la \phi | 
\bar{s} \gmmu s \ro \la 
K | \bar{s} \gmmu b | B \ra |^2} = \frac{1}{3} \left(\frac{
f_{\eta'}}{ f_{\phi}} \right)^2  \simeq 2.5 \cdot 10^{-2} 
\eeq
where we have used the definition $ \la \phi | 
\bar{s} \gmmu s \ro =  \varepsilon_{\mu} f_{\phi} 
m_{\phi} $ with $ f_{\phi} \simeq 240 \; MeV $ known 
experimentally from the $ \phi \rightarrow e^{+} e^{-} $
decay ( $ \varepsilon_{\mu} $ stands 
for the polarization vector of the $ \phi$-meson). As $ 
Br(B \rightarrow 
K \phi) \simeq 1 \cdot 10^{-5} $ \cite{Desh}
, we obtain a very small magnitude $ Br(B \rightarrow 
K \eta') \simeq 2.5 \cdot 10^{-7} $ in reasonable agreement with 
(\ref{7}). It is now obvious that corrections due to a 
non-factorizability of penguin operators, magnetic 
penguins contributions\footnote{It has been 
argued that the magnetic penguin 
operator $ O_{11} $ enhances the branching ratio for the $ 
b \rightarrow s \phi $ decay by 20-30 \% \cite{Desh2}. We expect
a similar (or, anyway, not
larger) effect of this operator for the $ B \rightarrow K \eta'
$ decay.},
 as well as  
$ O(1/m_b , 1/N) $ terms which have been  omitted in (\ref{6}),
cannot substantially change the estimate (\ref{7}). 
We therefore conclude that the image of the $ \eta' $ meson 
as the
SU(3) singlet quark state, made exclusively 
of the $ u,d,s $ quarks, is not adequate 
to the problem at hand. To avoid possible misunderstanding,
we should note that the axial anomaly is in fact taken into 
account in the above mechanism. However, its role there is 
merely to fix the residue of the quark currrent (\ref{5}) into
the $ \eta'$. 

\section{The gluon mechanism in $ B \rightarrow K \eta' $}

Here we suggest an alternative mechanism for the $ B \rightarrow
 K \eta' $ decay 
which is based on the well known fact 
that the $ \eta' $ 
is a very special meson strongly coupled to the gluons.  
Therefore, the process of interest can be 
mediated by the $ b \rightarrow c $ 
decay followed by a conversion of the $c$-quarks into
 gluons.  
This means that the matrix element 
\beq
\label{9}
\lo \bar{c} \gmmu \gmf c | \eta'(p) \ra = i \fc p_{\mu}
\eeq
is non-zero due to the $ c \rightarrow  gluons $ transitions. 
Of course,
since one deals here with virtual c-quarks, this matrix 
element is 
suppressed by the $ 1/m_{c}^2 $ factor. On the other hand,
 the c-quark is not very heavy and, taken together with
 the Cabbibo
enhancement of the $ b \rightarrow c $ transition in
 comparison
to $ b \rightarrow u $, the suggested scenario of the 
$ B \rightarrow K \eta' $  can be
expected to successfully compete
with the standard one described in Sect.2.
Actually, this gluon mechanism will be argued to dominate 
the decay. 
To get a feeling of how large the residue (\ref{9}) 
must be in order to explain the data (\ref{Kim}), we 
reverse the arguments and estimate this quantity
``experimentally" under assumption that the 
proposed gluon mechanism exhausts
the $ B \rightarrow K \eta' $ decay. A corresponding 
number is easy to calculate. In the factorization 
approximation the amplitude takes the form
\beq
\label{a}
M = \frac{G_{F}}{\sqrt{2}} V_{cb} V_{cs}^* a_1 
\la \eta'(p)| \bar{c} \gmmu \gmf c \ro \la K(q) |
\bar{s} \gmmu b | B(p+q) \ra 
\eeq
(here $ a_1 \simeq 0.25 $ , see Sect.2). For the $ B 
\rightarrow K $ transition form factor (\ref{4}) we use
the dipole formula  
\beq
\label{b}
f_{+} (p^2) = \frac{f_{+}(0)}{ 1 - p^2/m_{\ast}^2 }
\eeq
with $ f_{+}(0) \simeq 0.32 \; , \; m_{\ast} \simeq 5 \; GeV
$ \cite{Cher,Bel}. Calculating now the decay width, we obtain
numerically the branching ratio in terms of 
 the residue $ \fc $ defined in (\ref{9}) 
\beq
\label{c}
Br( B \rightarrow K \eta') \simeq 3.92 \cdot 10^{-3} \cdot
\left( \frac{f_{\eta'}^{(c)}}{1 \; GeV} \right)^2
\eeq
which together with the data (\ref{Kim}) implies 
the ``experimental" value (we use the central value 
of the branching ratio (\ref{Kim}) )  
\beq
\label{exp}
\fc \simeq 140 \; MeV \; \; (``exp")
\eeq 
This number may seem to be too large for the proposed 
mechanism to work 
as it is only a few times smaller than the
analogously normalized residue 
$ \lo \bar{c} \gmmu \gmf c | \eta_c (p)
\ra = i f_{\eta_c} p_{\mu} $ with
$  f_{\eta_c} \simeq 400 \; MeV $ known from the
$ \eta_c \rightarrow 2 \gamma $ decay. 
However, as will be argued in the rest
of this paper, the theory is able to produce such a 
large residue $ \fc $.  
In effect, the 
gluon mechanism completely overplays the
standard one
by two orders of magnitude in the decay width,
and reconciles a theoretical prediction for the 
$ B \rightarrow K \eta' $ decay with the data.

We now proceed to a theoretical calculation of the $ \eta' $ 
residue of the 
charmed axial current (\ref{9}). Making use of the anomaly 
equation, we 
obtain from (\ref{9}) 
\beq
\label{10}
\fc  = \frac{1}{m_{\eta'}^2} \lo 2 m_c \bar{c} i \gmf c +
\atop \er 
\eeq
Since the $c$-quark is heavier then the $\eta'$, it cannot
contribute the matrix element (\ref{10}) on a valence level.
It does, however, contribute when propagating in a loop. 
The $c$-quark in the loop is subject to external
gluon fields populating the $ \eta'$. A technical
tool which allows to evaluate the corresponding contribution 
to the matrix element (\ref{10}) is the Operator Product
Expansion in inverse powers of the $c-$quark mass (the heavy 
quark expansion):   
\beq
\label{11}
2 m_c \bar{c} i \gmf c = - \atop - \frac{1}{16 \pi^2
 m_{c}^2 }
g^3 f^{abc} G_{\mu \nu}^a \tilde{G}_{\nu \alpha}^b 
G_{\alpha \mu}^c 
+ \ldots 
\eeq
A detailed derivation of Eq.(\ref{11}) is given in
 Appendix, while 
here we restrict ourselves by a few comments. The first
 term in the 
right hand side of (\ref{11}) is the usual anomaly 
term with 
the opposite sign. This sign can be easily understood if
 one reminds
that the anomaly term corresponds to a subtraction
 of the 
Pauli-Villars regulator from the naive divergence $ 2
 m_q \bar{q}
i \gmf q $ of the axial current $ \bar{q} \gmmu 
\gmf q $. On the 
other hand, the Pauli-Villars contribution is a special
 case of 
the heavy quark expansion (\ref{11})
with the strict limit $ M_{R} = \infty $. 
The cancellation of leading terms $ \sim G \tilde{G}
$ which do not depend on $ m_c $ is in agreement 
with the intuitive idea that heavy quarks cannot 
contribute matrix elements over light hadrons\footnote{ A well 
known example is the problem of
a light particle mass: 
 in the conformal anomaly equation
\[
m_{\eta'}^2 = \el \theta_{\mu \mu} \er = 
\el \sum_{q} m_q \bar{q} q \er + \frac{ \beta(g^2)}{2g} 
\el G_{\mu \nu}^2 \er 
\]
(where $ \theta_{\mu \mu} $ is the trace of the 
momentum-
energy tensor and the sum is taken over all quark flavors)
the heavy quark can only contribute when propagating in
 a loop,
and its contribution cancels a corresponding contribution
 of the 
heavy quark to the $ \beta$-function in the second term 
.Thus, the $c$-quark does not contribute the $ \eta' $ mass
in the limit $ m_c \rightarrow \infty $.}
 in the limit 
$ m_{Q} \rightarrow \infty $.
The second term in (\ref{11}) is the gluon operator
 of the 
lowest (after the anomaly term) dimension $ d = 6 $. 
We omit one
 more $ d= 6 $ 
operator $ (D_{\mu} G_{\mu \nu})(D_{\alpha}G_{\alpha \nu} 
) $ which can 
be related to a four-quark operator using the equation
 of motion. 
One can show that a contribution of this operator is
 suppressed 
both parametrically ( in $ N $ and $ \alpha_s $) and
 numerically. Dots
in (\ref{11}) stand for higher dimensional operators 
which we do not 
address here. To this accuracy we therefore obtain
\beq 
\label{12}
\fc = - \frac{1}{16 \pi^2 m_{\eta'}^2 } \frac{1}{m_{c}^2}
\lo g^3 f^{abc} G_{\mu \nu}^a \tilde{G}_{\nu \alpha}^b 
G_{\alpha \mu}^c \er
\eeq
We note that the gluon operator in (\ref{12}) correspond 
to a normalization point $ \mu^2 \simeq  m_c^2 $.

We have thus reduced the problem to a calculation of the 
matrix element of the purely gluonic operator
(\ref{12}). Apart from a trivial rescaling of the 
normalization point, this matrix element is essentially 
defined by a low energy physics on a scale $ \sim 1 \; GeV
$. One could therefore think that we did not make any
progress at all as matrix elements of gluon operators
are usually not easy to calculate. A situation with 
the $ \eta'$ is, however, exceptional, and the matrix
element (\ref{12}) is amenable to a theoretical study. 
We will now show that it can be evaluated 
 making use     
of the 
large $ N $ line of reasoning 
 along with the property $ m_{\eta'}^2 
\sim 1/N $ 
in close analogy with a way Witten has addressed a very
similar matrix element $ \lo G \tilde{G} \er $.
The fact that $  m_{\eta'}^2 \sim 1/N $ was established 
by  Witten
\cite{Witten} a long time ago in connection to the 
celebrated $ U(1) $ problem.
Witten's objective was to understand within the large N 
approach 
how massless quarks are able to bring the correlation
 function
of the topological density \beq 
\label{14}
T (p) = i \int dx e^{ipx} \lo T \{ \atop(x) \; \atop(0) \}
 \ro
\eeq
 at zero momentum $ p \rightarrow 0 $
down to zero 
as required by the chiral anomaly, if quark loops 
are suppressed by a power of $ N $, and thus apparently
 do not 
show up to leading order in $ 1/N $. To lowest order 
in $ 1/N $ the two-point function (\ref{14}) is given
 by sums
over one-hadron intermediate states
\beq
\label{15}
T(p) = \left[ \sum_{glueballs} \frac{ a_{n}^2}{ M_{n}^2 -
 p^2} + subtractions \right] 
+ \frac{1}{N} \sum_{mesons} \frac{ c_{n}^2}{ m_{n}^2 - p^2}  
\eeq
where $ M_{n} ,  a_n$ and $ m_{n} , 
N^{-1/2} c_n $ stand for the masses and residues of $n$th
 glueball
and meson states, respectively. Here $ a_n, c_n = O(N^0) $,
and moreover to lowest order in  $ 1/N $ the  
glueball residues $ a_n $ do not depend on whether 
massless quarks are present in the Lagrangian or not. 
Therefore 
the first term in (\ref{15}) to leading order in
 $ 1/N $ 
corresponds to the two-point function in pure 
Yang-Mills
theory (gluodynamics). The crucial observation made by
 Witten 
\cite{Witten} and Veneziano \cite{Ven} was 
that while a cancellation of the two terms in (\ref{15}) is 
not possible at generic $ p^2 \neq 0 $, it can happen at
 $ p^2 = 0 $
, if there is a single meson with $ m^2 \sim 1/N $,
which would then cancel the whole sum over glueballs together
with subtraction terms in 
(\ref{15}). Witten and Veneziano have further 
identified this meson with the $ \eta' $ since the
 latter is the 
lightest flavor singlet preudoscalar state in nature. 
  
 The reason we have repeated at length the argument 
due to Witten and Veneziano is 
that, as is easy to see, it carries practically without
 a 
word of alteration over any non-diagonal 
correlation function of the
 topological
density $ \atop $ and arbitrary gluon operator with
 the $ O^{- +}   
$ quantum numbers. Choosing for such the three gluon 
operator defining
the matrix element (\ref{12}), we obtain the 
relation
\bea
\label{16}
\lo  \gcub \er \frac{1}{m_{\eta'}^2} \el \atop \ro
=  \nonumber \\
- i \int dx \lo T \{ \gcub(x) \; \atop (0) \} \ro_{YM}
 + O(\frac{1}{N})
\eea
where the subscript $ YM $ means that the correlation
 function 
in (\ref{16}) refers to pure Yang-Mills theory. 
Its calculation 
will be addressed in the next section. Here we would like
to mention that, as follows from (\ref{16}), the residue
of interest $ \lo g^3 G \tilde{G} G \er = O(N^{-1/2}) $.

\section{QCD low energy theorems}

The two-point function (\ref{16}) is a new unknown 
quantity which 
has to be evaluated in order to estimate the matrix 
element (\ref{9}). 
Analogously to the diagonal correlation function of the 
topological density (\ref{14}), it vanishes in the 
perturbation 
theory to all orders in $ \alpha_s $, since the 
topological density
is a total derivative whose matrix elements are all 
zero at the 
perturbative level. Therefore the correlation function
 (\ref{16}) can
only be non-zero due to non-perturbative contributions.
The goal of this section is to estimate (\ref{16}).

The idea of how to study correlation functions of gluonic 
currents with $ O^{+(-)} $ quantum number (like our Eq.(\ref{16})
) was suggested long ago \cite{NSVZ}, and we would like to 
shortly repeat it here.  
It has been known
for a long time that in channels with scalar or
 pseudoscalar quantum 
numbers leading contributions to correlation functions 
of composite
operators are related not to standard 
vacuum condensates, but rather to the so-called direct instantons
\cite{NSVZ}. The motivation to introduce this object into the
theory was a very strong indication that the standard QCD
sum rules \cite{SVZ} with standard power corrections due to 
local vacuum condensates are not able to describe the $ 0^{+(-)}
$ channels. In different words, OPE does not reproduce there
a scale of phenomena which is 
exactly known from elsewhere. A source of this effect 
was found in existence of direct instantons.

A meaning of this object is best explained if one 
considers first a two-point function of (pseudo-)
scalar gluon currents at large Euclidean momentum 
$ Q^2 $. In this case a leading non-perturbative contribution
is obtained when the momentum $ Q $ is transfered as a 
whole to a second vertex by a vacuum field (this is allowed
by quantum numbers of the current) which therefore must be 
of small size $ \rho \sim 1/Q $. Such situation corresponds
to a small coupling regime, in which the quasiclassical 
approximation becomes accurate. The vacuum field is therefore
classical; it is the famous BPST instanton 
\cite{Pol}. Thus, a recipe for
calculation of the direct instanton contribution at large $ Q^2 $
is simple: the gluon field in the current must be substituted 
by the instanton. The integral over the instanton size is then 
dominated by small $ \rho \sim 1/Q $. However, with going down 
to a resonance region $ Q^2 \sim a \; few \; GeV^2 $, this 
simple picture breaks down \cite{NSVZ} - instantons start to 
interact strongly with each other and large size vacuum fields, 
and the one instanton (or, what is the same, dilute instanton
gas) approximation stops making sense. 
A consistent calculation of instanton contribution in this
case becomes     
a complicated problem which requires going beyond 
the dilute instanton gas approximation and taking into 
account 
instanton interactions e.g. in a form suggested by the 
instanton liquid 
vacuum model of Shuryak and Diakonov-Petrov (see  
 \cite{Shur,Shur2} and references therein). We shall 
not proceed with this approach which is basically 
a specific model of the QCD vacuum. Rather, we will 
 follow an
alternative method which was 
proposed by Novikov et al. (NSVZ)
\cite{NSVZ}.
It makes use of a strong assumption that though  a vacuum 
field transferring a small momentum $ Q $ resembles only a 
little the original undeformed BPST instanton, it nevertheless
retains the (anti-) self-duality property of the latter 
\beq
\label{17}
G_{\mu \nu}^a = \pm \tilde{G}_{\mu \nu}^a
\eeq
in absence of massless quarks, i.e. in YM theory (where 
instanton
transitions are not suppressed by fermion zero modes).
 This conjecture has been supported by explicit calculations 
of next-to-leading (after direct instantons) non-perturbative 
corrections for two-point functions of the scalar $ G^2 $ and 
pseudoscalar $ G \tilde{G} $ currents at moderate $ Q^2 $. Up 
to an overall sign, they turn out the same (the direct instanton
contributions are identical in both channels by definition). 
Therefore, a further extrapolation of the self-duality 
selection rule (\ref{17}) to even lower $ Q^2 \simeq 0 $ is 
expected to be correct at least to a 100 \% accuracy.
In fact, a  phenomenologically successful 
mass formula for
 the $ \eta' $ derived in Ref.\cite{NSVZ} by using the 
selection rule (\ref{17}) implies that an actual accuracy
of this approximations is of the order of 50 \%. 
This mass formula has been obtained\footnote{Somewhat 
differently from Witten's arguments,
this was done without an explicit reference to the large $ N $
picture,
but rather using the fact that the $ \eta' $ is light on a 
characteristic mass scale in the $ 0^- $ channel, and 
therefore its inclusion is a local $ Q^2 \simeq 0 $ effect
in the momentum space which must nullify the two-point
function (\ref{14}) in full QCD.}
by relating 
the residue of the $ \eta' $ to the value of 
the two-point function of topological density in YM
 theory at $ Q^2 = 0 $, while the latter was evaluated using
the selection rule (\ref{17}) and a low energy theorem (see 
Eq.(\ref{18}) below). As a result, $ m_{\eta'}^2  $  was 
found proportional 
to the gluon condensate $ \la g^2 G^2 \ra $ in  
YM theory.

Following the same logic, 
we therefore assume that the self-duality selection rule
(\ref{17}) 
can be also
applied to the correlation function of interest (\ref{16}). 
If this is the case, the value of the latter is fixed by the 
low energy theorem \cite{NSVZ}
\beq
\label{18}
i \int dx \lo T \{ O (x) \; \frac{\alpha_s}{
4 \pi} G^2 (0) 
\} \ro_{YM} = \frac{2 d }{b} \la O \ra_{YM}
\eeq
Here $ O(x) $ is arbitrary color singlet local
operator of canonical dimension $ d $ made of 
gluons and $ b = 11/3 N $ stands for the 
first coefficient of the $ \beta$-function
in pure Yang-Mills theory. As a derivation of the 
fundamental relation (\ref{18}) is rather simple, for the 
sake of completeness we would like to remind it here.
One starts with a redefinition of the gluon 
field
\beq
\label{19}
\bar{G}_{\mu \nu} \equiv g_0 G_{\mu \nu}
\eeq
where $ g_0 $ is the bare coupling constant of QCD 
defined at the cut-off scale $ M_0 $. Then the path
integral representation immediately yields the relation
\beq
\label{20}
i \int dx \lo T \{ O (x) \; \bar{G}^2 (0) 
\} \ro = - \frac{d}{ d(1/4 g_0^2)} \la O \ra
\eeq
On the other hand, the renormalizability and the dimensional
transmutation phenomenon in a massless theory (either QCD with 
massless quarks or gluodynamics) ensure that
\beq
\label{21}
\la O \ra = const \; \left[ M_0 \exp \left(- \frac{8 \pi^2}{b
g_0^2} \right) \right]^d 
\eeq
with the choice $ b = 11/3 N -2/3 n_f $ (where $ n_f $ is
a number of flavors) or $ b =11/3 N $, respectively.
Finally, performing the differentiation yields 
the low energy theorem 
(\ref{18}). More accurate derivation (which 
gives the same final result) including a regularization
of ultra-violet divergences in (\ref{18}) can be found 
in \cite{NSVZ}. Note that by definition perturbative 
contributions are always subtracted in vacuum condensates 
like $ \la O \ra $. Using now the low energy theorem (\ref{18}) 
for the particular choice $ O = g^3 G^3 $, we obtain
\beq
\label{22}
\fc \simeq \frac{3}{4 \pi^2 b} \frac{1}{m_c^2} \frac{
\la g^3 G^3 \ra _{YM}}{ \lo \atop \er } 
\eeq
This is the main result of this section.
Coming back to our original definition (\ref{9}), we 
see that we have related the residue of the charmed 
axial current into the $ \eta'$ with apparently 
completely unrelated quantity which is the value of 
cubic gluon condensate in pure Yang-Mills theory
(the matrix element of the topological density is
known \cite{NSVZ}, $ \lo (\alpha_s/ 4 \pi) G \tilde{G}
\er \simeq 0.04 \; GeV^3 $).
This object will be addressed below, while here 
we would like to end up this section with a 
discussion of one important conceptual point. It is well
known that usually non-diagonal transitions between quarks and
gluons or between quarks of different flavors are 
suppressed - this is the 
famous Zweig rule. A most popular theoretical explanation
of this phenomenon exploits the large N argumentation: in this 
limit  all non-diagonal in flavors two-point functions are 
suppressed by 
powers of $ 1/N $ relatively to diagonal ones. Thus, at
$ N \rightarrow \infty $ any mixing dies off. On the other 
hand, the Zweig rule is strongly violated in the scalar and 
pseudoscalar $ 0^{ + (-)} $ channels. A well known example 
of this violation is provided by the pseudoscalar meson nonet:
while in the vector channel the $ \rho- $ and $ \omega-$ 
mesons are almost pure
mass eigenstates of the broken flavor SU(3) $ 
\rho \sim (\bar{u}
u - \bar{d} d )$ , $ \omega \sim \bar{s} s $ and
 the $ \rho -
\omega $ 
mixing is small, nothing similar is observed in the 
pseudoscalar nonet.
There the $ \eta $ is predominantly the octet $ \eta 
\sim (\bar{u}u 
+ \bar{d} d - 2 \bar{s} s )$, which means that the mixing 
is 100 \%.  
The theoretical explanation  \cite{NSVZ} as to 
why the Zweig 
rule is violated 
for quark or gluon currents with the  $ 0^{+(-)} $ 
quantum numbers is that 
in these channels there are direct instantons which are
 able
to convert quarks into gluons and vice versa $ \bar{q} q 
\leftrightarrow g g $ at the 
classical level without any suppression. Literally speaking,
all factors ensuring a smallness drop out: powers of the 
coupling constant disappear since the instanton field is
strong, $ G_{\mu 
\nu } \sim 1/g $, and geometrical loop factors like $ 1/(16 \pi^2)
$ do not arise because there are no loops. One of the striking
examples of such kind is a conversion of gluons into photons
\cite{NSVZ}: while naively the amplitude
\beq
\label{23}
\lo - \frac{b \alpha_s}{ 8 \pi} G^2 | 2 \gamma \ra
= O \left( \left( \frac{\alpha_s}{\pi} \right)^2 \frac{\alpha}{\pi}
\right)  
\eeq
in fact it is only $ O( \alpha/\pi) $ according to a strict low
energy theorem \cite{NSVZ} which reads
\beq 
\label{low}
\lo \frac{\beta(\alpha_s)}{4 \alpha_s} G^2 | \gamma(k_1)
\gamma (k_2) \ra = \frac{\alpha}{3 \pi} N n_f \la Q_q^2
\ra F_{\mu \nu}^{(1)}  F_{\mu \nu}^{(2)}
\eeq
( Here $ \la Q_q^2
\ra $ is the mean quark electric charge and 
$ F_{\mu \nu}^{(i)} = k_{\mu}^{(i)} \varepsilon_{\nu}^{(i)}
-  k_{\nu}^{(i)} \varepsilon_{\mu}^{(i)} $ , $ i = 1,2 $
stands for the fields strength of a plane wave.)
 We would like to note that 
this process is rather similar to our case.  
 The gluons proceed to photons through a loop of the $c$-quark,
but the perturbatively expected suppression factor
$ (\alpha_s/\pi)^2 $ does not occur!

A natural question to ask now is the following: does all this  
mean that the large N picture is strongly violated in the 
$ J^P = 0^{+(-)} $ channels ? A related question is
whether the experimental fact that
$ m_{\eta'}^2/m_{\rho}^2 > 1 $ despite that $ m_{\eta'}^2 \sim
1/N $ , $ m_{\rho}^2 = O( N^0) $ implies such a violation.
 The answer to both questions is no. 
Moreover, neither the direct instantons, nor
the low energy theorems are at variance 
with the large N picture. 
The point is that, as was explained in 
Ref.\cite{NSVZ}, $ N $ is the dimensionless parameter, and 
true mass relations look rather like $ m_{\eta'}^2 = M^2 /N $,
where $ M^2 $ is some mass scale. It is usually tacitly 
assumed within the large N reasoning that this mass scale is 
universal for all hadrons. That this is not the case was 
demonstrated by NSVZ \cite{NSVZ}: the mass scale $ M^2 $ is
not universal but determined by quantum numbers of 
a channel considered. More concretely, this mass 
is set by a scale at which the asymptotic freedom is violated
in the particular channel. The latter depends drastically 
on whether
direct instantons are allowed (which is the case in the $
 0^{+(-)}
$ channels) or not. If they are there, an interaction of 
external current
with vacuum fields is very strong, and the asymptotic freedom 
breaks down at very small distances, i.e. a characteristic
 mass scale
in this channel is not the typical hadron mass $ \sim
 m_{\rho}^2 $,
but rather much higher. It is therefore clear that the
 second of 
the questions posed above is not properly formulated:
the mass of the $ \eta' $ should be compared not with $ 
 m_{\rho}^2 $, but to a characteristic mass in the $ O^- $
 channel
which is $ \sim 15 \; GeV^2 $ \cite{NSVZ}. This fact is
 the 
``experimental" evidence that the $ 1/N $ argumentation is
 quite 
accurate for the $ \eta' $, and $ 1/N $ terms, omitted in 
(\ref{16}),
are small {\it in comparison to both explicitly written 
terms}.
At the same time, the above consideration explains why the 
effects
which we discuss are not negligibly small
numerically: despite of the fact
that the matrix element $ \lo g^3 G \tilde{G} G \er  \sim 
N^{-1/2}
$, a large dimensional parameter in front of $  N^{-1/2}
 $ is 
able to make it large in reality.

\section{How large is $ \la g^3 G^3 \ra $ in pure 
gluodynamics ?}

We are now returning to the mainstream of our consideration. 
Our task has reduced, according to Eq.(\ref{22}), to a 
determination of the cubic gluon condensate $  
\la g^3 G^3 \ra_{YM} $ in pure YM theory. Note that 
this quantity does not have to (and in fact does not) 
coincide with the cubic condensate $\la g^3 G^3 \ra$
in the real world. While for the latter there exists 
a semi-phenomenological estimate \cite{Zhit1}
\beq
\label{24}
\la g^3 G^3 \ra = ( 0.06 - 0.1) \; GeV^6
\eeq
obtained within the QCD sum rules approach, it is of no
direct use in Eq.(\ref{22}). Unfortunately, we are not aware
of any method (except, probably, the lattice
approach) which could
reliably calculate $ \la g^3 G^3 \ra_{YM} $ to an accuracy
of, say, 20 \%. Because of this uncertainty, we are unable
to get a theoretical prediction for $ Br(B \rightarrow
K \eta') $ with precision comparable to the experimental
one. What instead will be argued in this section is that 
different estimates of the value $  \la g^3 G^3 \ra_{YM} $
enable one to claim that
the large number (\ref{Kim}) is within the realms of our current
understanding of {\it non-perturbative} QCD. We believe
that this statement is interesting by itself in 
view of a failure of the standard approach to 
this problem (see Sect.2). It will be shown below 
that $ \la g^3 G^3 \ra_{YM} > \la g^3 G^3 \ra_{QCD} $ 
(this result follows directly from
the theory and has a status of theorem), 
and moreover numerically
\beq
\label{25}
 \la g^3 G^3 \ra_{YM} = (0.4 - 1.4) \; GeV^6
\eeq
The first tool we are going to use is again the low 
energy theorem (\ref{18}). Let us recall how a very similar
question on a value of the condensate $ 
\la g^2 G^2 \ra_{YM}
$ was addressed in the classical paper \cite{NSVZ}. First
of all, we note that this condensate corresponds to an 
imaginary world in which all quarks are very heavy. This 
world could be obtained from the real one when the masses 
of light $ u-, d- $ and $ s-$quarks are smoothly drawn up
to some large value $ m_q $. By the decoupling theorem,
this mass must not be very large - once it reaches the 
confinement scale $ \mu \simeq 200 \; MeV $, the heavy quarks
decouple and do not influence any more the gluon condensate.
One the other hand, the
confinement scale $ \mu $ is not too far from the 
$ s-$quark mass $ m_s \simeq 150 \; MeV $. Therefore, one
can expect that a value of $ \la g^2 G^2 \ra $ in a world 
with $ m_u = m_d = m_s \equiv m_q \simeq 150 \; MeV $ 
would give a reasonable estimate for the YM condensate
$ \la g^2 G^2 \ra_{YM}
$. In the linear approximation in $ m_u , m_d $ we need
to know the derivative
\beq
\label{26}
\frac{d}{d m_q} \la \frac{\alpha_s}{\pi} G^2 \ra
= - i \int dx \; \lo T \left\{ \frac{\alpha_s}{\pi}
G^2 (0) \; \bar{q} q (x) \right\} \ro
\eeq
The latter two-point function is fixed by the
low energy theorem (\ref{18}) to be proportional
to the quark 
condensate known from elsewhere. 
Since the quark condensate is negative,
it follows that $ (d/dm_q) \la g^2 G^2 
\ra > 0 $. In this way NSVZ have 
obtained an estimate 
\beq
\label{27}
\la \frac{\alpha_s}{\pi} G^2 \ra_{YM} = 
(2-3) \la \frac{\alpha_s}{\pi} G^2 \ra_{QCD}    
\eeq
As has been argued in \cite{NSVZ}, the sign of effect 
$ \la g^2 G^2 \ra_{YM} >
 \la g^2 G^2 \ra_{QCD} $ is in perfect
agreement with the instanton picture. Indeed, raising
the quark masses diminishes the chiral suppression of 
instantons and therefore increases $ \la G^2 \ra $. 

Proceeding analogously, we write
\beq
\label{28} 
\frac{d}{dm_q} \la g^3 G^3 \ra = 
- i \int dx \; \lo T 
\left\{ g^3 G^3 (0) \; \bar{q}q(x) \right\} \ro
\eeq
The difference from the case of quadratic gluon condensate
(\ref{26}) is that the two-point function does not coincide
with the low energy theorem as it was in Eq.(\ref{26}).
Therefore, its value is not known exactly. Nevertheless,
the low energy theorem (\ref{18}) can still be used to 
{\it estimate} the correlation function (\ref{28}). 
To this end, consider the relation (\ref{18}) 
in QCD with $ b = 11/3 N - 2/3 n_f $ for 
three different operators
\bea
\label{29}
i \int dx \; \lo T \left\{ \frac{\alpha_s}{4 \pi} G^2
(x) \;  \frac{\alpha_s}{4 \pi} G^2 (0) \right\} \ro
&=& \frac{2}{b} \la \frac{\alpha_s}{\pi} G^2 \ra 
\nonumber \\
i \int dx \; \lo T \left\{ \bar{q} q 
(x) \;  \frac{\alpha_s}{4 \pi} G^2 (0) \right\} \ro
&=& \frac{6}{b} \la \bar{q} q  \ra   \\ 
i \int dx \; \lo T \left\{ g^3 G^3
(x) \;  \frac{\alpha_s}{4 \pi} G^2 (0) \right\} \ro
&=& \frac{12}{b} \la  g^3 G^3 \ra  
\nonumber 
\eea
We now assume that these low energy theorems are 
saturated by some effective glueball state $ 
\sigma $ with 
$ m_{\sigma} \sim 1 \; GeV $. 
It should be stressed that we do not
insist on existence of a real 
narrow glueball resonance with such
mass. Actually, introducing such a 
(fictitious?) glueball amounts to an 
effective description of the physics of $ 
0^+ $ channel. Analogous methods have 
been used in a similar in spirit problem
of a strange content of the nucleon 
\cite{Zhit2} which also deals with the 
$ O^+ $ channel. Note that the 
glueball mass drops out in the 
final result (\ref{32}). Introducing the residues
\beq
\label{30}
\lo \frac{\alpha_s}{4 \pi} G^2
| \sigma \ra = \lambda_1 \; , \; \lo \bar{q}q | \sigma
\ra = \lambda_2 \; , \; \lo g^3 G^3 | \sigma \ra = \lambda_3
\eeq
we put Eqs.(\ref{29}) in the form
\bea
\label{31}
\frac{\lambda_1^2}{m_{\sigma}^2} & =& \frac{2}{b} 
\la \frac{\alpha_s}{\pi} G^2 \ra \nonumber \\
\frac{\lambda_1 \lambda_2}{m_{\sigma}^2} &= &\frac{6}{b} 
\la \bar{q}q \ra  \\
\frac{ \lambda_3 \lambda_1}{m_{\sigma}^2} &= &\frac{12}{b} 
\la g^3 G^3 \ra \nonumber 
\eea
Using these equations and assuming the same scalar 
glueball dominance in (\ref{28}), we obtain
\beq
\label{32}
\frac{d}{dm_q} \la g^3 G^3 \ra = 
- \frac{\lambda_3 \lambda_2}{m_{\sigma}^2} = 
- \frac{36}{b} \frac{ \la \bar{q}q \ra \la g^3 G^3 
\ra }{ \la \frac{\alpha_s}{\pi} G^2 \ra }
\eeq
As $ \la \bar{q} q \ra < 0 $, the sign of the derivative 
is fixed: $ (d/dm_q) \la g^3 G^3 \ra > 0 $. Moreover,
while an expected accuracy of the estimate (\ref{32})
is of the order of 100 \%, the sign is entirely model
independent and in fact is fixed by the positivity 
of spectral densities in Eqs.(\ref{29}). By the 
same token as for $(d/dm_q) \la g^2 G^2 \ra $, this 
sign agrees with expectations based on the instanton
picture of the vacuum. Numerically, using the 
values\footnote{The number for the quark condensate corresponds
to the normalization point $ 1 \; GeV^2 $. Here we would like 
to point out that the normalization pont in our Eq.(\ref{22})
is $ \mu^2 \simeq m_c^2 $, while the low energy 
theorems refer, strictly speaking, to much lower $ 
\mu \simeq 500 \; MeV $. In view of a large numerical 
uncertainty of our 
results we neglect this perturbative evolution. Still, 
one has
to bear in mind that the large anomalous dimension of the 
$ G^3 $ operator $ \gamma = -18 $ \cite{Mor} is working 
in the same direction:
taking it into account is only able to enhance our final result
(\ref{25}) or the estimate (\ref{33}) by a factor of two.} 
$ \la \bar{q}q \ra \simeq - 0.017 \; GeV^3 ,
\la (\alpha_s/ \pi ) G^2 \ra \simeq 0.012 \; GeV^4
$, in the linear approximation in $ m_q $ Eq.(\ref{32})
yields for $ m_q = 150 \; MeV $
\beq
\label{33}
 \la g^3 G^3 \ra_{YM} \simeq 4.4 \la g^3 G^3 \ra_{QCD}
= (0.26 - 0.44) \; GeV^6
\eeq
where we have used the estimate (\ref{24}). Note 
that the effect of going from QCD to gluodynamics 
is larger for the cubic condensate (\ref{33}) than 
for the quadratic one (\ref{27}). This fact is
a direct consequence of the low energy theorem (\ref{29}),
and can also be  
easily understood on dimensional grounds: 
since $ \la g^3 G^3 \ra \sim \la g^2 G^2 \ra^{3/2} $,
increasing $ \la g^2 G^2 \ra $ by a factor $ \sim 
2.5 $ (see Eq.(\ref{27})) yields raising of $ \la 
g^3 G^3 \ra $ by $ (2.5)^{3/2} \simeq 4 $ in comparison 
to its value in QCD. Eq.(\ref{33}) constitutes 
our first estimate for a value of the cubic gluon 
condensate in pure YM theory which is based on 
the low energy theorems (\ref{29}) and the 
semi-phenomenological information on its value in real
world (\ref{24}). 

Another possible source of information on the value 
$ \la g^3 G^3 \ra_{YM} $ is the instanton liquid vacuum
model (see \cite{Shur,Shur2}). Two basic parameters of this 
model are average instanton size $ \rho_c $ and  
inter-instanton separation $ R $. The latter 
parameter is chosen such that to reproduce the 
phenomenological value of the gluon condensate
$ \la g^2 G^2 \ra $ using the fact that each instanton
contributes a fixed amount
$ \int dx \; g^2 G^2 = 32 \pi^2 $ to this quantity. This 
yields the number $ R \simeq 1/200 \; MeV^{-1} $
for the phenomenological value (in QCD) $
\la (\alpha_s/\pi) G^2 \ra \simeq 0.012 \; GeV^4 $.
This number provides the upper limit for
the instanton density in QCD as it implies 
that the entire gluon condensate is due to instantons. 
On the other hand, the ratio $ \rho_c/R $ does not 
depend on the value of condensate and is fixed dynamically
to be $ \rho_c / R \simeq 1/3 $. A value of the cubic 
condensate calculated in this model was found to be 
essentially larger than the semi-phenomenological
number (\ref{24}): 
\beq
\label{34}
\frac{ \la g^3 G^3 \ra }{ \la g^2 G^2 \ra}
= \frac{12}{ 5 \rho_c^2} \simeq 0.9 \; GeV^2 \; , \; 
\la g^3 G^3 \ra \simeq 0.4 \;GeV^6 
\eeq
(the formula for $ \la g^3 G^3 \ra $ in terms of 
the instanton radius $ \rho_c $ was first established 
in \cite{SVZ}). We would like to make the following 
comment in reference to the result (\ref{34}). For 
our purposes it is more suitable to discuss the
instanton vacuum picture not in QCD, but in  
pure YM theory. We note that the instanton vacuum 
is more simple in gluodynamics than in QCD because of absence 
of the chiral suppression of instantons. In this 
case the inter-instanton separation must be chosen 
to fit the gluon condensate $ \la g^2 G^2 \ra $ 
in YM theory, which is according to (\ref{27})
larger than the corresponding number in QCD. On the 
other hand, for 
gluodynamics the ratio $ \rho_c /R \simeq 1/3 $ remains
the same. Using (\ref{34}) and the estimate (\ref{27}), 
we obtain  
\beq
\label{35}
\la g^3 G^3 \ra_{YM} \simeq 1.7 \; GeV^6
\eeq
which is a few times larger than our first estimate 
(\ref{33}). We feel that this result provides 
the upper estimate for the quantity of interest, 
and the true answer for 
$ \la g^3 G^3 \ra $ lies somewhere in between of the 
two numbers (\ref{33}) and (\ref{35}). 

Finally, we would like to discuss information 
on vacuum condensates, which is available from lattice simulations.
The quadratic gluon condensate in YM theory on the 
lattice was reported to be 
\[
\la \frac{\alpha_s}{\pi} G^2 \ra = \left\{ 
\begin{array}{ll} 0.15 \; GeV^4  & , \; SU(2) 
\; \cite{Al} \\
0.10 \; GeV^4  & , \; SU(3) \; \cite{Giac}
\end{array} \right. \]
We do not feel qualified enough to discuss a precision of 
these calculations. In particular, it is not very clear 
(at least to us) whether the large scale separation
is accurately performed to make possible a comparison
with the SVZ definition. Still, it seems undoubtful that 
these results point in the same direction as (\ref{27}):
the quadratic gluon condensate in YM theory is 
essentially larger than in QCD. As for the cubic condensate,
though a corresponding Monte Carlo data does exist, it 
is rather difficult to extract from it a value of the 
non-perturbative part of $ \la g^3 G^3 \ra $ \cite{G3}.
 An estimate of this quantity
can still be obtained in an indirect way as the lattice 
simulations of instantons suggest\footnote{E.V. Shuryak, 
private communication. See
also \cite{Shur2}.}
 that the average size 
of instantons in gluodynamics 
on the lattice is approximately $ 1/400 \; MeV $ which is 
a bit larger than the value predicted by the instanton liquid model. 
In this case Eq.(\ref{34}) together with 
the above value of the quadratic condensate 
yields an estimate (for the 
SU(3) color group)
\beq
\label{36}
\la g^3 G^3 \ra_{YM} \simeq 1.5 \; GeV^6 
\eeq
which is numerically close to (\ref{35}). Comparing  
finally all three estimates (\ref{33}),(\ref{35}) and 
(\ref{36}),
we suggest that while (\ref{33}) presumably gives a lower bound
for the number of interest, (\ref{35}) and (\ref{36}) seem 
to set up an upper limit with a possible short distance enhancement.
A reasonable compromise yields our final estimate given above by 
Eq.(\ref{25}).

 We are now in a position 
to estimate the principal input in (\ref{a}) which is the 
residue $ \fc $ of the charmed axial current into the $ \eta' $,
and which was the main object of our consideration in this
paper. Using (\ref{22}) and (\ref{25}), we obtain the following 
answer for this parameter:
\beq
\label{37}
\fc  = ( 50 - 180) \; MeV
\eeq
Note that literally the ``experimental" number (\ref{exp})
corresponds to the value of the condensate $ 
\la g^3 G^3 \ra_{YM}
 \simeq 1 \; GeV^6 $ which is about a midpoint of our 
prediction (\ref{25}).
Given the accuracy of our result (\ref{37}), we thus conclude
that the gluon mechanism seems to be sufficient 
to describe the data (\ref{Kim}).
Unfortunately, we are currently unable to improve 
 our estimate (\ref{37}), where the main source 
of uncertainty is due to a poor knowledge of the 
cubic condensate in YM theory (\ref{25}). Some ways
to do this will be discussed in the next section.

\section{Conclusions}

In this paper we proposed a theory of the 
$ B \rightarrow K \eta' $ decay. We showed that at the 
quark level this process proceeds via the $ b \rightarrow \bar{c}
c s $ weak decay followed by a conversion of the $c$-quark pair
directly into the $ \eta' $ which is possible due to 
a presence of a non-valence Zweig rule violating ``intrinsic
charm" component of the $ \eta' $ wave function. We have found 
that a mechanism of breaking down the Zweig rule in our 
case is of a purely non-perturbative origin. We have further
evaluated a most important ingredient of the 
factorized $ B \rightarrow K \eta' $
amplitude, which is the matrix element of the charmed current
$ \lo \bar{c} \gmmu \gmf c \er $, using a combination of the 
Operator Product Expansion technique, large $ N $ approach and
QCD low energy theorems. Our results demonstrate that the 
proposed mechanism is likely to exhaust an extremely large 
branching ratio measured by the CLEO collaboration, with a certain 
reservation for a poor accuracy of our final answer (\ref{37}).
We do not pretend to have made a 
numerically reliable calculation of the matrix
element $ \lo \bar{c} \gmmu \gmf c 
| \eta'(p) \ra = i \fc p_{\mu} $ which
in our opinion determines the  $ B 
\rightarrow K \eta' $ decay width.
We have rather 
presented a semi-quantitative (but parametrically
well established) picture which demonstrates a
close relation between this matrix element  and 
properties of the vacuum of QCD (or, more accurately,  
YM theory). Though
the obtained result looks unexpectedly large,
it has a good explanation within QCD, and is related to
 strong 
fluctuations in vacuum $0^{+(-)}$ channels.
The main source of uncertainty in our approach is  
well localized and related to a value of the cubic condensate 
$ \la g^3 G^3 \ra_{YM} $ which is currently not known with enough 
precision. 
 Yet, 
the  exact low energy theorems indicate that this matrix element
is quite large. The instanton vacuum model and lattice calculations 
also seem to favor a large value of this condensate. In this 
reference, more refined calculations of 
this quantity are highly welcome. In particular, it would 
be very interesting if this condensate 
could be reliably extracted from lattice simulations. 
An alternative way which can be suggested to improve 
the determination of the condensate $ \la g^3 G^3 \ra_{YM} $ is 
akin to the idea of the QCD sum rule approach \cite{SVZ}. If we
had other physical processes which essentially depend on the 
same cubic condensate, it could be then fixed ``phenomenologically"
once and forever with an expected consistency between theoretical
predictions for different physical amplitudes. We plan to return 
to these issues elsewhere. 

We would like to emphasize that the conclusion on a large
Zweig rule violating $c$-quark component of the $ \eta' $
certainly goes beyond the particular example of the 
$ B \rightarrow K \eta' $ decay and may well be important
in other physical processes. We repeat that there are two
basic reasons for a large magnitude of the residue $ \fc $:
(1) the $c$-quark mass is not too far from the hadron 
scale $ 1 \; GeV$ and (2) the Zweig rule is badly broken 
down in the $ 0^{+(-)} $ channels. While there is nothing 
particularly special (except for its numerical effect) about
the first factor, the second one is specific to the unique 
nature and quantum numbers of the $ \eta' $. Therefore, we 
do not expect that any other than $ \eta' $ light particle
could yield a similar contribution to the $ B $ decay.
In a more general content, there is an increasing evidence
for importance of non-valence Zweig rule violating
components in hadrons. We remind that the problem
of the strange quarks in the nucleon (the so-called 
$\pi N \; 
 \sigma- $term) is resolved  
\cite{Zhit2} (see also \cite{EKar}) within 
physics which is very similar to
that discussed in this paper.    
Furthermore, there are many other examples where
``intrinsic" non-valence configurations (including, in 
particular,
the ``intrinsic charm" hadron components) seem important 
(see e.g. a recent review \cite{Brodsky} and \cite{BrKar}).
All these examples unambiguously demonstrate that non-valence
components of hadron can be sizeable. In QCD terms such a 
situation means that a corresponding matrix element has 
a non-perturbative origin without the naively expected 
$ \alpha_s/\pi $ suppression. We have shown in this paper 
that this experimentally testable physics is amenable 
to a theoretical control. In this respect, the $ \eta' $ 
from $ B \rightarrow \eta' $ decays is an 
excellent laboratory for a study of fundamental
properties of strongly interacting QCD.

\section*{Acknowledgments}

We are grateful to P. Kim for interesting discussions 
during his visit to UBC, which 
initiated this study. We would like to thank
A. DiGiacomo and E.V. Shuryak for correspondence. 
\clearpage
\appendix
\def\theequation{\thesection.\arabic{equation}}

\section*{Appendix}

\def\thesection{A}
\setcounter{equation}{0}
 
The purpose of this Appendix is to derive Eq.(\ref{11}) in the 
text. A  convenient machinery for such class of problems was 
invented by Schwinger a long time ago
\cite{Sch}. The Schwinger technique
allows in many instances (e.g. when one is interested in a short
distance expansion) to operate with a propagator in external 
field without a specification of the field. A result has
a form of expansion in powers of the field and its derivatives.
The relevant object in our problem is the $c$-quark propagator
at short distances $ \sim 1/m_c $, and the expansion in the 
external gluon field amounts to a representation of the 
propagator in a form of series in powers of $ \Lambda_{QCD}/m_c $
which is a sort of OPE. We refer the interested reader to a 
pedagogical technical review \cite{rev} for more detail 
and relevant references.

The Schwinger operator approach is based on a realization 
of commutation relations of the coordinate  $ X_{\mu}
$ and momentum $ P_{\mu} $ ($ P_{\mu} = i D_{\mu} $ , where 
$ D_{\mu} $  is the covariant derivative) operators 
\bea
\label{a1}
\left[ P_{\mu}, X_{\nu} \right] &=& i g_{\mu \nu} \nonumber \\
\left[ P_{\mu}, P_{\nu} \right] &=& i g G_{\mu \nu}^a T^a 
\eea
where $ T^a $ are the generators of the color group. One 
introduces in the 
coordinate space a formal complete set of states $ |x \ra $ 
as the eigenstates of the coordinate operator $ X_{\mu} $ :
\bea
\label{a2}
X_{\mu} |x \ra &=& x_{\mu} |x \ra \nonumber \\
\la y | x \ra &=& \delta (x-y) \\
 \int dx |x \ra \la x | &=& 1 \nonumber  
\eea
while in this basis the momentum operator $ P_{\mu} $ acts 
as the covariant derivative
\beq
\label{a3}
\la y| P_{\mu} | x \ra = \left( i \frac{\partial}{\partial 
x_{\mu}} + g A_{\mu}^a (x) T^a \right) \delta(x-y)
\eeq
In these notations we have to evaluate the expression
\beq
\label{a4}
\bar{c} i \gmf c = \la x | Tr \left\{ \gmf ( P \! \! \! \!/ - 
m)^{-1} \right\} | x \ra
\eeq
where $ m $ is the $c$-quark mass and
$ Tr $ denotes the trace over both color and Lorentz 
indices. Using a resolution of unity 
\[ 
1 = ( P \! \! \! \! / + m)^{-1} ( P \! \! \! \!/ + m)   \]   
and the formula ( $ \sigma_{\mu \nu} = i/2 ( \gmmu \gmnu - 
\gmnu \gmmu) $ )
\[ 
 \pdir^2 = P^2 + \frac{1}{2}\sigma_{\mu \nu}  G_{\mu \nu}^a T^a 
\; , \] 
we expand (\ref{a4}) in powers of $ \sigma G \equiv 
\sigma_{\mu \nu}  G_{\mu \nu}^a T^a $ 
\bea
\label{a5}
\bar{c} i \gmf c &=& m \la x | Tr  \left\{ \gmf
\frac{ g^2}{4} \frac{1}{(P^2 - m^2)^2} (\sigma G )  
\frac{1}{P^2 - m^2} (\sigma G ) 
  \right. \nonumber \\
&-& \left. \frac{g^3}{8} \frac{1}{(P^2 - m^2)^2} (\sigma G )
\frac{1}{P^2 - m^2} (\sigma G )
\frac{1}{P^2 - m^2} (\sigma G ) + \cdots \right\} | x \ra
\\
&\equiv& D_1 + D_2 + \ldots \nonumber 
\eea
where dots stand for higher power terms in the external 
field expansion. It is convenient to start the calculation
with the second term in (\ref{a5}). It explicitly contains 
the gluon field $ G_{\mu \nu} $ to the third power, and 
therefore to our accuracy one can neglect the 
non-commutativity of the operators $ P_{\alpha} $ and $
G_{\mu \nu} $. Thus
\beq
\label{a6}
D_2 = - \frac{g^3}{8} m Tr \left\{ \frac{1}{(P^2 - m^2)^4} 
\gmf ( \sigma G)^3 \right\} 
\eeq
To the same accuracy the momentum operator $ P_{\mu }$
in the denominator of (\ref{a6}) is substituted by the c-number
$ p_{\mu} $, and the $x$-integration is performing using 
the formula 
\[ 
\la x | ( P^2 - m^2 )_{A=0}^{-(n-1)} | x \ra 
= \int \frac{d^4 p}{(2 \pi)^4 } \; (p^2 - m^2 )^{-(n-1)} 
= \frac{ (-1)^n m^{6-2 n} }{ 16 \pi^2 i (n-2) (n-3) }
\]
which yields
\beq
\label{a7}
D_2 = - \frac{ i g^3}{ 2^8 \cdot 3 \cdot \pi^2 \cdot m^3}
Tr \left\{ \gmf ( \sigma G )^3 \right\} 
\eeq
A calculation of the trace over the Lorentz indices gives 
\[ 
Tr \left\{ \gmf ( \sigma G )^3 \right\} = -2^5 
\cdot tr_c G_{\mu \nu}
\tilde{G}_{\nu \alpha} G_{\alpha \mu}  \] 
where $ tr_c $ stands for the trace over the color indices. 
The latter is 
\[ 
tr_c  G_{\mu \nu}
\tilde{G}_{\nu \alpha} G_{\alpha \mu} = \frac{i}{2} f^{abc}
 G_{\mu \nu}^a 
\tilde{G}_{\nu \alpha}^b  G_{\alpha \mu}^c \equiv 
\frac{i}{2} G \tilde{G} G  \] 
and finally we obtain
\beq 
\label{a8}
D_2 = - \frac{1}{ 16 \cdot 3 \cdot \pi^2 \cdot m^3} 
g^3 G \tilde{G} G 
\eeq
A calculation of the first term in (\ref{a5}) is more tedious as
now we have to take into account the non-commutativity of 
operators in order to evaluate this expression to the $
  G \tilde{G} G $ accuracy.
Using the identity
\bea
\label{id}
(\sigma G) \frac{1}{P^2 - m^2} &=&   \frac{1}{P^2 - m^2}(P^2 - m^2)
( \sigma G )  \frac{1}{P^2 - m^2} \nonumber \\
&=&  \frac{1}{P^2 - m^2} 
( \sigma G) +  \frac{1}{P^2 - m^2} [ P^2, (\sigma G) ] 
 \frac{1}{P^2 - m^2} 
\eea 
we write
\bea
\label{a9}
D_1 &=& \frac{g^2}{4} m \la x | Tr \left\{ \gmf 
 \frac{1}{(P^2 - m^2)^3} (\sigma G)^2 + \gmf  
\frac{1}{(P^2 - m^2)^3} [ P^2, ( \sigma G) ]  
\frac{1}{P^2 - m^2} ( \sigma G ) \right\} | x \ra \nonumber \\
& \equiv & \Pi_{1} + \Pi_2 
\eea
The first term in (\ref{a9}) is readily calculated to yield
\beq
\label{a10}
\Pi_1 = - \frac{g^2} { 32 \cdot \pi^2 \cdot m}  G_{\mu \nu}
\tilde{G}_{\mu \nu} \; , 
\eeq
while the second one needs more care. Let us perform first the 
trace over the Lorentz indices:
\beq
\label{a11}
\Pi_2 = - 2 i g^2 \la x | Tr \left\{  \frac{1}{(P^2 - m^2)^3}
[ P^2 , G_{\mu \nu} ]  \frac{1}{P^2 - m^2} \tilde{G}_{\mu \nu}
\right\} | x \ra
\eeq
Calculating the commutator 
\[ 
[ P^2 , G_{\mu \nu} ] = P_{\lambda} [ P_{\lambda}, G_{\mu \nu}
] + [ P_{\lambda}, G_{\mu \nu} ] P_{\lambda} = 
2 i P_{\lambda} ( D_{\lambda} G_{\mu \nu}) + D^2 G_{\mu \nu}
\] 
we obtain
\bea
\label{a12}
\Pi_2 &=& 4 g^2 m \la x | Tr \left\{  \frac{1}{(P^2 - m^2)^3}
 P_{\lambda} ( D_{\lambda} G_{\mu \nu})  \frac{1}{P^2 - m^2}
\tilde{G}_{\mu \nu} \right\} | x \ra \nonumber \\
&-& 2 i g^2 m \la x | Tr \left\{  \frac{1}{(P^2 - m^2)^3}
 (D^2  G_{\mu \nu})  \frac{1}{P^2 - m^2}
\tilde{G}_{\mu \nu} \right\} | x \ra \nonumber \\
&=&  4 g^2 m \la x | Tr \left\{  \frac{1}{(P^2 - m^2)^4}
 P_{\lambda} ( D_{\lambda} G_{\mu \nu})  
\tilde{G}_{\mu \nu} \right\} | x \ra \nonumber \\
&+&  4 g^2 m \la x | Tr \left\{  \frac{1}{(P^2 - m^2)^4}
[ P^2,  P_{\lambda} ( D_{\lambda} G_{\mu \nu}) ] 
 \frac{1}{P^2 - m^2}
\tilde{G}_{\mu \nu} \right\} | x \ra \nonumber \\
&-&  2 i g^2 m \la x | Tr \left\{  \frac{1}{(P^2 - m^2)^4}
 (D^2  G_{\mu \nu}) 
\tilde{G}_{\mu \nu} \right\} | x \ra + \ldots 
\eea
One can easily see that the first term in this expression 
gives rise to higher dimensional operators and thus does not 
contribute to our accuracy. However, the second term does contain
a needed power of the gluon field since
\[ 
[P^2 , P_{\lambda} ( D_{\lambda} G_{\mu \nu}) ] = 
2 i P_{\lambda} P_{\beta} D_{\beta} D_{\lambda} G_{\mu \nu}
+ \ldots = i \{   P_{\lambda}, P_{\beta} \} 
 D_{\beta} D_{\lambda} G_{\mu \nu} + \ldots 
\] 
which results in  
\beq
\label{a14}
\Pi_2 = 2 i g^2 m \la x | Tr \left\{  \frac{2}{(P^2 - m^2)^5}
 \{   P_{\lambda}, P_{\beta} \} 
 D_{\beta} D_{\lambda} G_{\mu \nu} \tilde{G}_{\mu \nu}
-  \frac{1}{(P^2 - m^2)^4}
 (D^2  G_{\mu \nu}) 
\tilde{G}_{\mu \nu} \right\} | x \ra 
\eeq
In the first term to our accuracy one can substitute 
$ P_{\mu} \rightarrow p_{\mu} $. We further use the Bianki
identity to evaluate
\beq
\label{a15}
 (D^2  G_{\mu \nu}) 
\tilde{G}_{\mu \nu} = D_{\alpha}( - D_{\nu} G_{\alpha \mu}
- D_{\mu} G_{\nu \alpha}) \tilde{G}_{\mu \nu} = 2 i 
G_{\alpha \mu } \tilde{G}_{\mu \nu} G_{\nu \alpha} 
\eeq
and eventually obtain 
\beq
\label{a16}
\Pi_2 = - \frac{1}{ 32 \cdot 3 \cdot \pi^2 \cdot m^3}
g^3 G \tilde{G} G 
\eeq
Finally, collecting together (\ref{a8}), ( \ref{a10}) and 
(\ref{a16}) and multiplying the whole answer by $ 2 m $, 
we arrive at 
\beq
\label{a17}
2 m \bar{c} i \gmf c = - \frac{\alpha_s}{ 4 \pi} G \tilde{G}
- \frac{1}{ 16 \pi^2 m^2 }g^3 G \tilde{G} G + \ldots
\eeq
which completes the proof of Eq.(\ref{11}).


\clearpage
\begin{thebibliography}{99}
\bibitem{CLEO} P. Kim [CLEO], Talk at FCNC 1997, Santa Monica,
CA (Feb 1997). \\
F. Wurthwein [CLEO], hep-ex/9706010.
\bibitem{Soni} D. Atwood and A. Soni, hep-ph/9704357.
\bibitem{Witten} E. Witten, Nucl. Phys. {\bf B156} (1979) 269.
\bibitem{Ven} G. Veneziano, Nucl. Phys. {\bf B159} (1979) 213.
\bibitem{Buras} A. Buras, M. Jamin, M. Lautenbacher and 
P. Weisz, Nucl. Phys. {\bf B400} (1993) 37. \\
 A. Buras, 
M. Jamin and M. Lautenbacher, Nucl. Phys. {\bf B400} (1993) 75. \\
M. Ciuchini, E. Franko, G. Martinelli and L. Reina, 
Nucl. Phys. {\bf B415} (1994) 403.
\bibitem{Desh} N. Deshpande and Xiao-Gang He, Phys. Lett. 
{\bf B336} (1994) 471. \\
 R. Fleischer, Z. Phys. {\bf C62} (1994) 81.
\bibitem{Ciu} M. Ciuchini, E. Franko, G. Martinelli, L. Reina
and L. Silvestrini, Phys. Lett. {\bf B316} (1993) 127.
\bibitem{NSVZ} V.A. Novikov, M.A. Shifman, A.I. Vainshtein
 and V.I. Zakharov,
 Nucl. Phys.
 {\bf B191} (1981) 301.
\bibitem{B-decays} M. Gourdin, A.N. Kamal, Y.Y. Keum and
X.Y. Pham, Phys. Lett. {\bf B333} (1994) 507. \\
T. Browder, K. Honscheid and S. Playfer, in B decays, ed. 
by S. Stone, World Scientific (1994).
\bibitem{Chau} L.L. Chau et. al., Phys. Rev. {\bf D43} (1991) 2176.
\bibitem{Desh2}  N. Deshpande, Xiao-Gang He and
J. Trampeti\' c, Phys. Lett. 
{\bf B377} (1996) 161.
\bibitem{Cher} V.L. Chernyak and I.R. Zhitnitsky, 
Nucl. Phys. {\bf B345} (1990) 137.
\bibitem{Bel}
 V.M. Belyaev, A. Khodjamirian and R. R\"uckl, 
Z. Phys. {\bf C60} (1993) 349.
\bibitem{SVZ} M.A. Shifman, A.I. Vainshtein and V.I. Zakharov,
 Nucl. Phys.
 {\bf B147} (1979) 385, 448, 519.
\bibitem{Shur} E.V. Shuryak, Nucl. Phys. {\bf B203} (1982) 93,
116,140. \\
D.I. Diakonov and V.Yu. Petrov, Nucl. Phys. {\bf B245} (1984) 259; 
{\bf B272} (1986) 457.\\
 E.V. Shuryak, Rev. Mod. Phys. {\bf 65} (1993) 1. 
\bibitem{Shur2} T. Schafer and E.V. Shuryak, hep-ph/9610451.
\bibitem{Pol} A.A. Belavin, A.M. Polyakov, A.S. Schwarz and
A.S. Tyupkin, Phys. Lett. {\bf 59B} (1975) 85.
\bibitem{Zhit1} A.R. Zhitnitsky, Sov. J. Nucl. Phys. 
{\bf 41} (1985)
805, 1035, 1331.
\bibitem{Zhit2} V.M. Khatsimovsky, I.B. Khriplovich and
A.R. Zhitnitsky, Z. Phys. {\bf C36} (1987) 455. \\
A.R. Zhitnitsky, Phys. Rev. {\bf D55} (1997) 3006.\\
A.R. Zhitnitsky, hep-ph/9611303.
\bibitem{Mor} A. Yu. Morozov, Sov. J. Nucl. Phys. 
{\bf 40} (1984) 505.
\bibitem{Al} B. Alles et al., Phys. Rev. {\bf D48} (1993) 2284.
\bibitem{Giac} M. Campostrini, A. DiGiacomo and 
Y. Gunduc, Phys. Lett. {\bf B225} (1989) 393.
\bibitem{G3} A. DiGiacomo, K. Fabricius and G. Paffuti,
Phys. Lett. {\bf 118B} (1982) 129. \\
 H. Panagopoulos and E. Vicari, Nucl. Phys. {\bf B332} (1990) 261.
\bibitem{EKar} J. Ellis, E. Gabathuler and M. Karliner,
Phys. Lett. {\bf B217} (1989) 173.
\bibitem{Brodsky} S. Brodsky, hep-ph/9609415, SLAC-PUB 7306.
\bibitem{BrKar} S. Brodsky and M. Karliner, hep-ph/9704379.
\bibitem{Sch} J. Schwinger, Phys. Rev. {\bf 82} (1951) 664.
\bibitem{rev} V. Novikov, M.A. Shifman, A.I. Vainshtein and 
V.I. Zakharov, Fortschr. Phys. {\bf 32} (1984) 585. 

\end{thebibliography}
\end{document}